\documentclass[fleqn,usenatbib]{mnras}

\usepackage{newtxtext,newtxmath}

\usepackage[T1]{fontenc}

\usepackage{graphicx}	
\usepackage{amsmath}	

\title[Crab Nebula detection with the 1st TAIGA IACT]{Detection of the Crab Nebula using a Random Forest Analysis of the first TAIGA IACT Data}

\author[Blank et al.]{
M. Blank,$^{1}$
M. Tluczykont,$^{1}$
A. Porelli,$^{2}$
R. Mirzoyan,$^{3}$
R. Wischnewski,$^{2}$
A. K. Awad,$^{1}$
M. Brueckner,$^{2,*}$
\\
$^{1}$Institut für Experimentalphysik, Universität Hamburg, Hamburg, D-22761 Germany\\
$^{2}$Deutsches Elektronen-Synchrotron DESY, Zeuthen, 15738 Germany\\
$^{3}$Max Planck Institute for Physics, Munich, 80805 Germany\\
$^{*}$Now at PSI, Z{\"u}rich, Switzerland\\
}

\date{Accepted XXX. Received YYY; in original form ZZZ}

\pubyear{2022}

\begin{document}
\newcommand{\ethreshold}{6\,TeV}
\newcommand{\signi}{8.5}
\newcommand{\non}{453}
\newcommand{\noff}{4053}
\newcommand{\excess}{163.5}
\newcommand{\phizero}{$(3.20\pm0.42)\cdot10^{-10}$\,TeV$^{-1}$cm$^{-2}$s$^{-1}$} 
\newcommand{\spectralindex}{$-2.74\pm0.16$}
\newcommand{\eref}{13\,\mathrm{TeV}}
\newcommand{\noffregions}{14}
\newcommand{\livetime}{$85\,\mathrm{h}$ }
\label{firstpage}
\pagerange{\pageref{firstpage}--\pageref{lastpage}}
\maketitle
\begin{abstract}
The Tunka Advanced Instrument for Gamma- and cosmic-ray Astronomy (TAIGA)
is a multicomponent experiment for the measurement of {TeV to PeV} gamma- and cosmic rays.
Our goal is to establish a novel hybrid {direct air shower} technique,
sufficient to access the energy domain of the long-sought Pevatrons.
{The hybrid air Cherenkov light detection technique combines the strengths of the HiSCORE shower front sampling array, and two $\thicksim$4\,m class,}
{$\thicksim$9.6$^\circ$ field of view Imaging Air Cherenkov Telescopes (IACTs). The HiSCORE array provides good angular and shower core position resolution, while the IACTs provide the image shape and orientation for gamma-hadron separation.
In future,}
{an additional muon detector will be used for hadron tagging at $\geq100$\,TeV energies.}
Here, only data from the first IACT of the TAIGA experiment are used.
A {\it random forest} algorithm was trained using Monte Carlo (MC) simulations and real data, and applied to \livetime of {selected} observational data tracking the Crab Nebula at a mean zenith angle of 33.5\,$^\circ$, resulting in a threshold energy of \ethreshold~for this dataset.
%
%
The analysis was performed using the gammapy package. A total of \excess~excess events were detected, with a statistical significance of \signi\,$\sigma$.
The observed spectrum of the Crab Nebula is best
fit with a power law
{above~\ethreshold}
with a flux normalisation of \phizero at a reference energy of $\eref$ and a spectral index of \spectralindex.
\end{abstract}

\begin{keywords}
gamma-rays: general -- instrumentation: detectors -- methods: observational
\end{keywords}



\section{Introduction}
Gamma-ray observations in the so far poorly explored energy range from ten to hundreds of TeV can ultimately solve the problem on the origin of Galactic cosmic rays up to knee energies.
TAIGA \citep{budnev2020} is located on the site of the Tunka-133 \citep{2012NIMPA.692...98B} cosmic ray detector (51°48'35''N, 103°04'02''E).  
{The experiment site is located in the Tunka valley, 50 km west of the southernmost tip of the Baikal lake in Siberia,}
at an altitude of 675\,m above sea level.
{Primary gamma and cosmic rays initiate extended air showers (EAS) in the atmosphere. The EAS is measured on the ground, using the air Cherenkov light detection technique as well as particle detectors.
TAIGA is implementing a novel hybrid method, combining the two existing approaches: the shower front sampling (HiSCORE) and the IACT techniques.
HiSCORE is an array with optical detector stations (35$^\circ$ half opening angle), which integrate Cherenkov light from EAS. Several IACTs (9.6$^\circ$ field of view) are placed at distances between 250\,m and 600\,m from each other.
In addition to the air-Cherenkov arrays, a scintillator-based charged particle detector array is being installed for muon tagging.}

The design of TAIGA is focused on the gamma-ray energy regime from a few TeV to several hundred TeV and cosmic ray observations above 100\,TeV.
%
The observed cosmic ray population up to PeV energies is considered to be fuelled by Pevatrons \citep{2007ApJ...665L.131G}, that accelerate cosmic rays up to 10$^{15}$\,eV (PeV) with sufficient power to balance the escape losses from the particles leaving the Galactic disk region.
These Pevatrons will necessarily be sources of gamma rays and possibly neutrinos with energies beyond 100\,TeV as a decay product of neutral and charged mesons (mainly pions) produced in inelastic scattering of the accelerated nuclei with the ambient medium \citep{2007ApJ...665L.131G}.
Neutrino observations by IceCube \citep{IceCube:2013cdw}, as well as recent
gamma-ray observations of Pevatrons by LHAASO \citep{cao:lhaaso:2021} have emphasised the need to explore this energy regime.
TAIGA aims to perform spectroscopic and morphological studies of Pevatrons.
%
In addition to gamma rays, the TAIGA data will be used to measure the chemical composition and directional anisotropy of cosmic rays in the transition regime from presumably Galactic (PeV) to extragalactic (several 100\,PeV) energies.
%
Finally, the science program of TAIGA includes topics in the field
of particle physics such as
the measurement of the proton-proton cross-section,
the search for quark-gluon plasma in air-showers,
photon-ALPs conversion in the Galactic magnetic field,
search for violation of Lorentz invariance \citep{2016CRPhy..17..632H},and search for heavy supersymmetric particles (wimpzillas) \citep{2017JCAP...05..036C}.

This article describes the analysis of Crab Nebula data taken during the commissioning of the first TAIGA-IACT using a {\it random forest} analysis method (see Section~\ref{sec:reco}).

Before this analysis is described and results are presented, the details of the detector components are introduced in the following Sections.


\section{TAIGA Detector components}
\label{sec:taiga}
TAIGA {consists of} three components.
{The first component, HiSCORE, samples the amplitude and arrival time of Cherenkov light emitted by the secondary particles of an EAS.}
The imaging air Cherenkov Telescopes (TAIGA-IACTs) {take snapshots of EAS developing in the atmosphere}; and finally, the scintillator-based TAIGA surface and underground detectors that provide a measurement of the {charged particle} component.

In total, 120 HiSCORE stations were deployed until recently, over an area of 1.1\,km$^2$. As of 2021, two 4\,m class IACTs with a 9.6$^\circ$ aperture are in operation,
and the third IACT is under completion.
%
The current Tunka Grande scintillator array consists of 19 scintillator stations deployed above and below the ground. In future, a larger TAIGA-Muon array \citep{Astapov2018} will be deployed to complement the air-Cherenkov components of TAIGA, for providing an efficient independent tool for gamma-hadron separation at energies above 100\,TeV.

\begin{figure}
\centering
\includegraphics[scale=0.25]{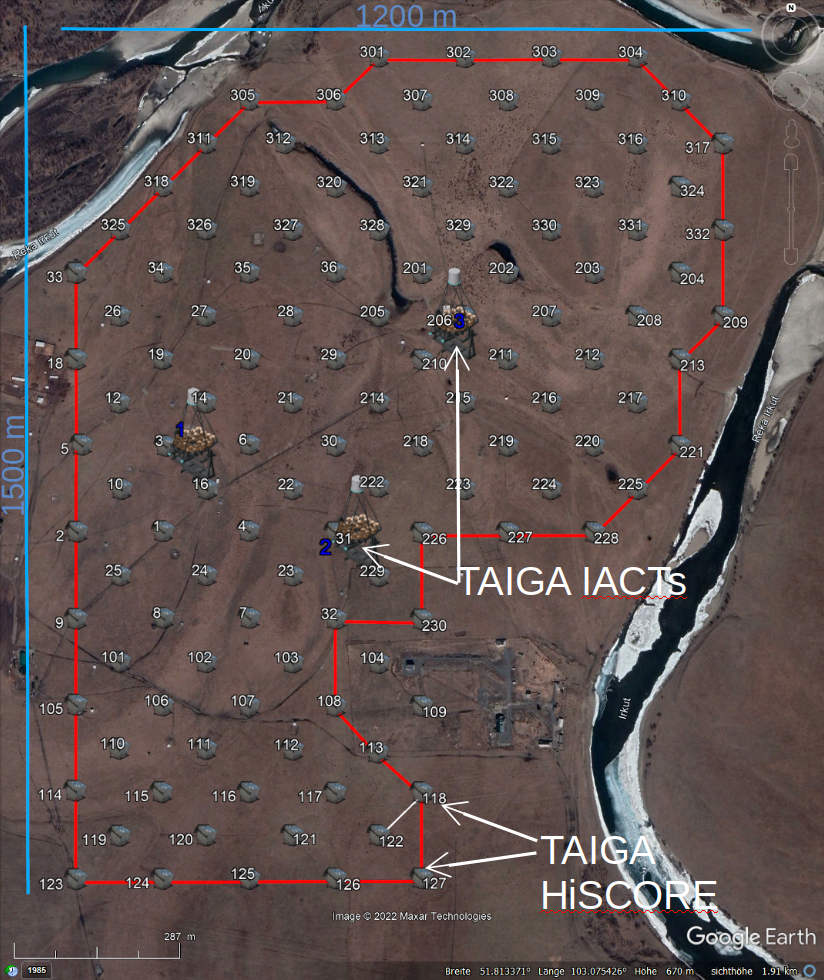}
\caption{\label{fig:layout}
Layout of the TAIGA-HiSCORE array with 3 IACTs. The first two IACTs (west and south) were deployed from 2016 to 2019. A third IACT (north-east) is under construction.
The background picture, obtained from Google-Earth, shows the location of TAIGA with the Irkut river flowing around the array. The drawn station and telescope sizes are not to scale and only used for the indication of their position. The currently active part of the array is indicated by the red perimeter (1.12\,km$^2$). Scale lengths are indicated in blue.
}
\end{figure}

The layout of the TAIGA-HiSCORE and TAIGA-IACT components is shown in Figure~\ref{fig:layout}.


The size of the TAIGA pilot-array should allow us to prove the principle of the hybrid method, as well as to deliver first astrophysical results by using this new technique.
In a future phase of TAIGA, a 10\,km$^2$ array with up to 1000 HiSCORE stations and 16 IACTs is planned. The spacing of the IACTs in the pilot array is closer than planned for the future setup, allowing to test the hybrid method against the classical stereoscopic approach. In the future array, the telescopes will be placed up to 600\,m apart from each other, thus maximising the total collection area for better coverage of the multi-TeV energy regime.

A recent overview of the TAIGA experiment and its detector components is given in \citep{budnev2020,kuzmichev2021}.

\section{The first IACT}
\label{sec:iact1}
The TAIGA IACTs follow a design similar to that of the former HEGRA telescopes 
\citep{Mirzoyan1994}.
The first TAIGA-IACT is a prototype design. The second telescope addresses some of the issues of the first and with the development of the third telescope, the design got frozen.

\subsection{Telescope Mechanics and Optics}
IACT 1 consists of a tessellated mirror dish of Davies-Cotton design with a diameter D of 4.3\,m and an F/D of 1.1. The dish is equipped with 34 mirrors of 60\,cm diameter each. In the first couple of years, the telescope was equipped with 29 round glass mirrors, of a thickness of 2\,cm, produced by the company Galaktika in Armenia. Additional 5 hexagonal mirrors were added later \citep{borodin:2020}.
for higher light collection, resulting in a total reflective area of 10\,m$^2$.
%
Each individual mirror facet is mounted on a mechanical system with one fixed joint and two actuators. These allow one to manually align the mirror facets {in order to achieve the best}
point-spread function (PSF) in the focus. We align the mirrors by using an artificial light source located at $\thicksim$800\,m distance, resulting in a PSF of less than 0.14$^\circ$.
%
The altitude and azimuth axes are driven by Phytron stepper motors and drive electronics. The position angle of each axis is controlled using 17-bit absolute shaft encoders. 
A CCD camera mounted on the reflector dish is monitoring the PMT camera and part of the sky. We use bright stars to measure the possible deviations and to correct the absolute pointing of the telescope.
The latter was evaluated to have an accuracy of  0.01$^\circ$ \citep{2019ICRC...36..833Z,Zhurov2021}. 
An additional test of the pointing accuracy was performed using measurements of the anode currents of the PMTs during a drifting passage of known stars through the pixels, confirming findings using the CCD camera results.
%
%
%
%
%

\subsection{Camera Design}
\label{sec:camera_design}
At the focal distance (4.75 m), a photomultiplier (PMT)-based camera with a field of view of 9.6$^\circ$ diameter is used to detect EAS images in Cherenkov light. The camera is based on Philips 560 XP1911 PMTs with a diameter of 19\,mm (effective photo-cathode diameter 15\,mm).

%
Winston cone-type light collectors are set on PMTs to increase the light collection efficiency of the camera and to reduce the albedo. The Winston cones are assembled in one structure and mounted on top of the PMT plane. The initial pilot design was far from optimal. The hexagonal-shaped walls of the cones {made of black plastic} have a wall thickness of 3\, mm. Unfortunately, the relatively large inter-pixel dead area {due to thick, non-transparent material} reduces the light collection efficiency by $\thicksim$30\,$\%$. An upgrade with higher-efficiency light cones is underway.
Each camera pixel covers an angular diameter of 0.36$^\circ$ in the sky.

The camera body walls are insulated against the cold winter temperatures\footnote{The temperatures reach values of typically -35$^\circ$\,C$^\circ$ during observational nights in winter.}
and the strong temperature gradients along its radius using a $\thicksim$10\,cm thick layer of Styrofoam. It is inserted between the outer skin of the camera and the actual aluminium body of the camera.
The heat released by the electronic components within the imaging
camera is circulated, heating the internal surface of the 15\,mm-thick
entrance window made from UV transparent acrylic glass. 
During daylight, a light-tight lid protects the front of the camera.
A hot air generator is installed on the rear side of the telescope. By
using a system of plastic tubes it blows hot air onto a concave side of
mushroom-like transparent reflectors. These are installed between every
three mirrors, ~10cm above their reflecting surface. The reflected heat
scatters away along the mirror surfaces, preventing these from dew
formation.
To prevent the connection cables from becoming stiff at very low temperatures,
these are covered by heating jackets.

\begin{figure}
    \centering
    \includegraphics[width=\columnwidth]{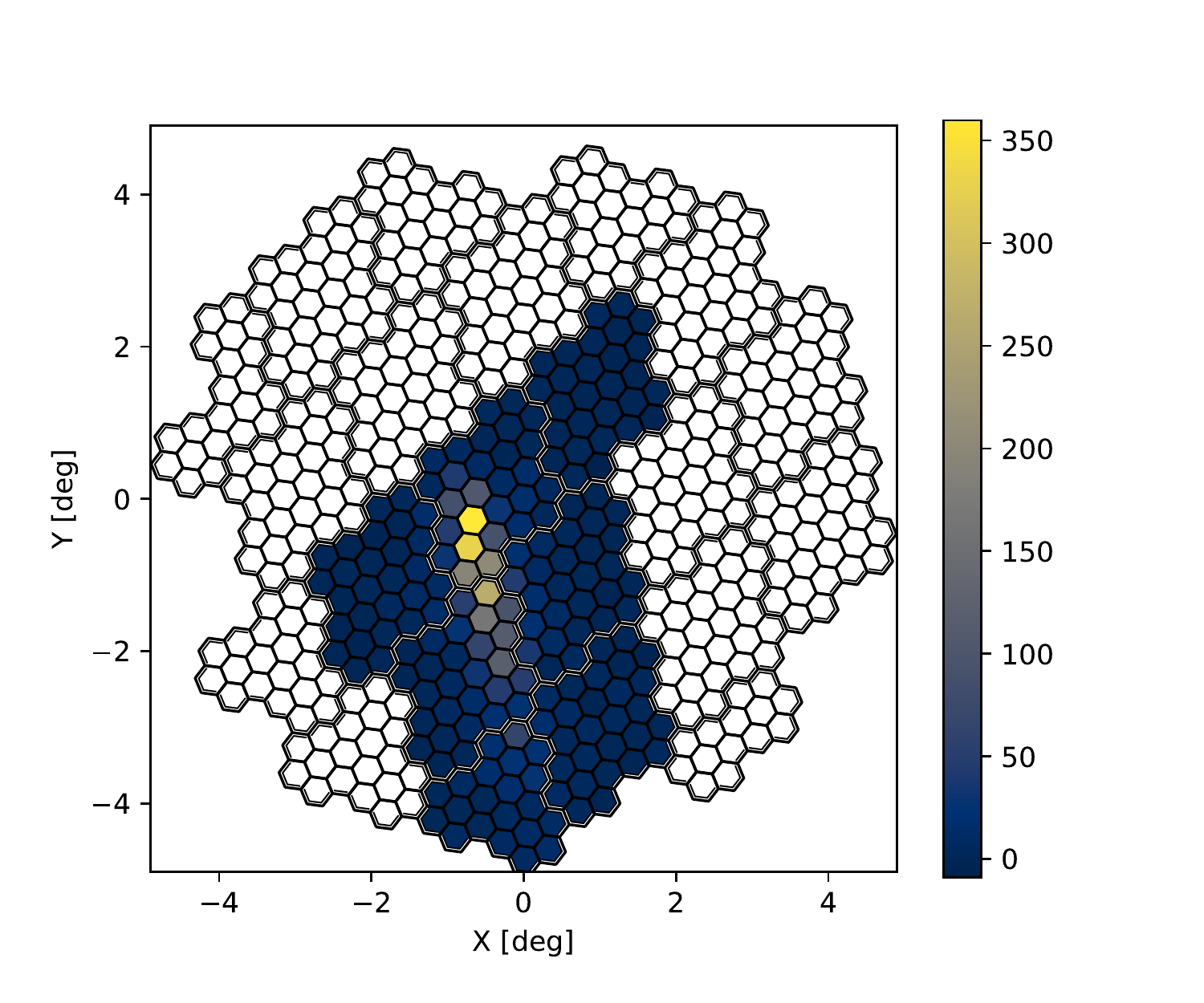}
    \caption{The layout of the PMT camera of IACT\,1. Cluster borders are shown in black double-lines. The color scale indicates the amplitude in photoelectrons for a typical event from real data. White PMT-pixels belong to clusters which are not read out.}
    \label{fig:camera-event}
\end{figure}

Groups of 7 neighbouring PMTs, selected to have similar gains, are connected to a common high-voltage (HV) power supply. Four groups of such 7 are organized into a cluster of 28 pixels (somewhat less at the camera edges). Figure~\ref{fig:camera-event} shows the cluster structure with a typical event from real data.
Each cluster is equipped with a readout board based on the 64-channel ASIC chip MAROC3 \citep{blin:2008}. For providing a wide dynamic range, two MAROC3 readout channels, each set to a different MAROC preamplifier gain, are used for reading out every single pixel in the camera. A fast shaper on the MAROC board is used for the trigger channel, and a somewhat slower one for the peak sensing readout.
%

After a pixel trigger is formed at time $t_{trg}$, the amplitude of the MAROC slow shaper is read out at time $t_{trg} + t_{hold}$. The hold time for the first IACT is set to 40 ns, locating the readout time on average 10\,ns after the slow-shaper peak.
The trigger decision is formed for each cluster separately, requesting at least 2 PMTs to produce an amplitude above a preset threshold.
In order to reduce the trigger rate due to light of the night sky (LoNS), a topological condition requiring the two pixels to be adjacent was introduced in 2019. {All data used in the present analysis were taken using this topological trigger condition.}
%
In order to limit the trigger rate,
the trigger-sensitive area of the camera is reduced to an inner radius of 3.7$^\circ$.
The cluster shape, elongated in one direction, and the trigger condition induce an inhomogeneity in the camera response at the trigger level, suppressing low energy events that lie along the border between two clusters. This issue is remedied by using a larger cut on the total image amplitude (size), at the cost of an increased energy threshold. During the first observation season, only triggered clusters were read out. This lead to a reduced reconstruction efficiency, due to incomplete images.
Later on, this trigger strategy has been improved by producing a trigger also in the neighbour cluster(s) if any of those pixels produced an amplitude in excess of 11 photoelectrons (p.e.).
As a further improvement, the second TAIGA IACT (not used here) was designed with a full camera readout.

Further details on the camera readout and slow control electronics can be found in \cite{Yashin2016,Lubsandorzhiev2019,budnev2020}.

%



\subsection{Reconstruction}
\label{sec:reco}
Real data and Monte Carlo (MC) simulations are reconstructed using the same algorithms. While MC simulation parameters are available directly in units of p.e., those for real data must first be calibrated and then converted into p.e. as described in Section~\ref{sec:cal}.

%

Two independent processing chains for the real data exist in TAIGA. The data processing used for obtaining the results in the present paper is described below.
Tail-cuts are used to clean the images from the noise induced by LoNS, requesting minimum amplitudes $a_1$ and $a_2$ for the brightest and second brightest neighbouring pixels.
Isolated pixels passing the threshold for the second brightest pixel, but without neighbour passing the first threshold, are rejected. The results presented here were obtained with tail-cuts $a_1 = 14$\,p.e. / $a_2 = 7$\,p.e..
The cleaned images are parameterized using the classical Hillas formalism \citep{1985ICRC....3..445H}.

%
%
%
For a stand-alone IACT, the direction of an event 
can be reconstructed by using the dependence of the elliptical shape of the image on the position of its center of gravity (COG) with respect to the source position in the camera. The parameter {\it disp}, depending on the ratio of the large to small axes of the ellipse, is used for that purpose.
An ambiguity on the impact point of the shower remains due to two possible positions of the direction on both sides of the image along its major axis (this is an artefact of using the second moments for the ellipse). This ambiguity can be resolved by using the third central moment (M3L).

In the analysis presented here, 
an algorithm based on classification trees trained by MC, {\it random forest} \citep{Breiman:2001hzm,Hengstebeck2007Measurement,2008NIMPA.588..424A}, 
is applied to the data. The input of the {\it random forest} are the altitude angle, the Hillas parameters {\it width, length, size, concentration}, the distance between COG and camera centre, the number of pixels, and M3L.
Based on the Python package {\it scikit-learn} \citep{scikit-learn}, two regressors were implemented to determine the parameter {\it disp} and the energy of the primary particle.
Furthermore, a random forest classifier was used to determine the nature of the primary particle.
The random forest training was carried out using {exclusively} MC simulations for gamma rays (see Section~\ref{sec:sim}), and real data for hadrons.
%
The output of the random forest classifier is a {\it gammaness} parameter that expresses the probability for the event to be an EAS initiated by a gamma-ray.
For gamma-hadron separation, a cut on {\it gammaness}$>$0.80 is applied. The cut value was optimised on the quality factor
\begin{equation}
    \label{eq:qfactor}
    Q = \frac{\epsilon_\gamma}{\sqrt{\epsilon_\mathrm{hadron}}}
\end{equation}
using gamma-ray MC simulations (weighted according to a Crab Nebula-like spectrum) and real data for hadrons. At a gammaness of 0.80, the gamma ray and hadron efficiencies are $\epsilon_\gamma = 0.47$ and $\epsilon_\mathrm{hadron} = 0.024$, yielding $Q=3.0$ at zenith distances of 30$^\circ$ to 40$^\circ$.
Furthermore, cuts are applied on the {\it leakage} ($<$0.15), {defined as the ratio of the sum of amplitudes of a given image at the camera edge to the total sum of amplitudes \citep{2002PhDT.......391S}} and the size ($>$80\,p.e.).
{The cut on the {\it leakage} is done to avoid miss-reconstruction from images truncated by the camera edge. The cut on the size is done to avoid using data in the sub-threshold range, below which the amplitude fluctuations of the IACT images are large. Another issue is that the above-mentioned inhomogeneity (Section~\ref{sec:camera_design}) in the camera response becomes important at small image sizes.}
%

%
%
%

In the next section, the MC simulations used for
training the {\it random forest} and for testing its performance are presented and compared to background from real data.
The results of the observation of the Crab Nebula data are described in Section~\ref{sec:results}.

A robust classical Hillas analysis, 
which is based on easy to understand 
principles,
was also implemented as a cross check.
%
%
%
%
%
More complex algorithms such as the {\it random forest} method used here
often provide better performances.

\subsection{Monte Carlo Simulation}
\label{sec:sim}
Two different Monte Carlo (MC) chains are used for the simulation of the TAIGA-IACT detector response. For air shower simulations, both chains rely on CORSIKA \citep{1998cmcc.book.....H}. One detector simulation is realised using the sim\_telarray package \citep{2008APh....30..149B}, with correspondingly adapted configuration files and additional processing for the simulation of the MAROC readout \citep{kunnas:phd,Blank2021},
and another one was specifically developed for TAIGA \citep{Postnikov2019}.
The former simulation chain is used here.
MC data are weighted according to a parametrisation of the differential spectra of cosmic ray primaries \citep{2003APh....19..193H} for nuclear charge Z$\ge$3, and a parametrisation from measurements by the DAMPE experiment for proton and He \citep{PhysRevLett.126.201102,2019SciA....5.3793A}. 

\begin{figure}
        \includegraphics[width=\columnwidth]{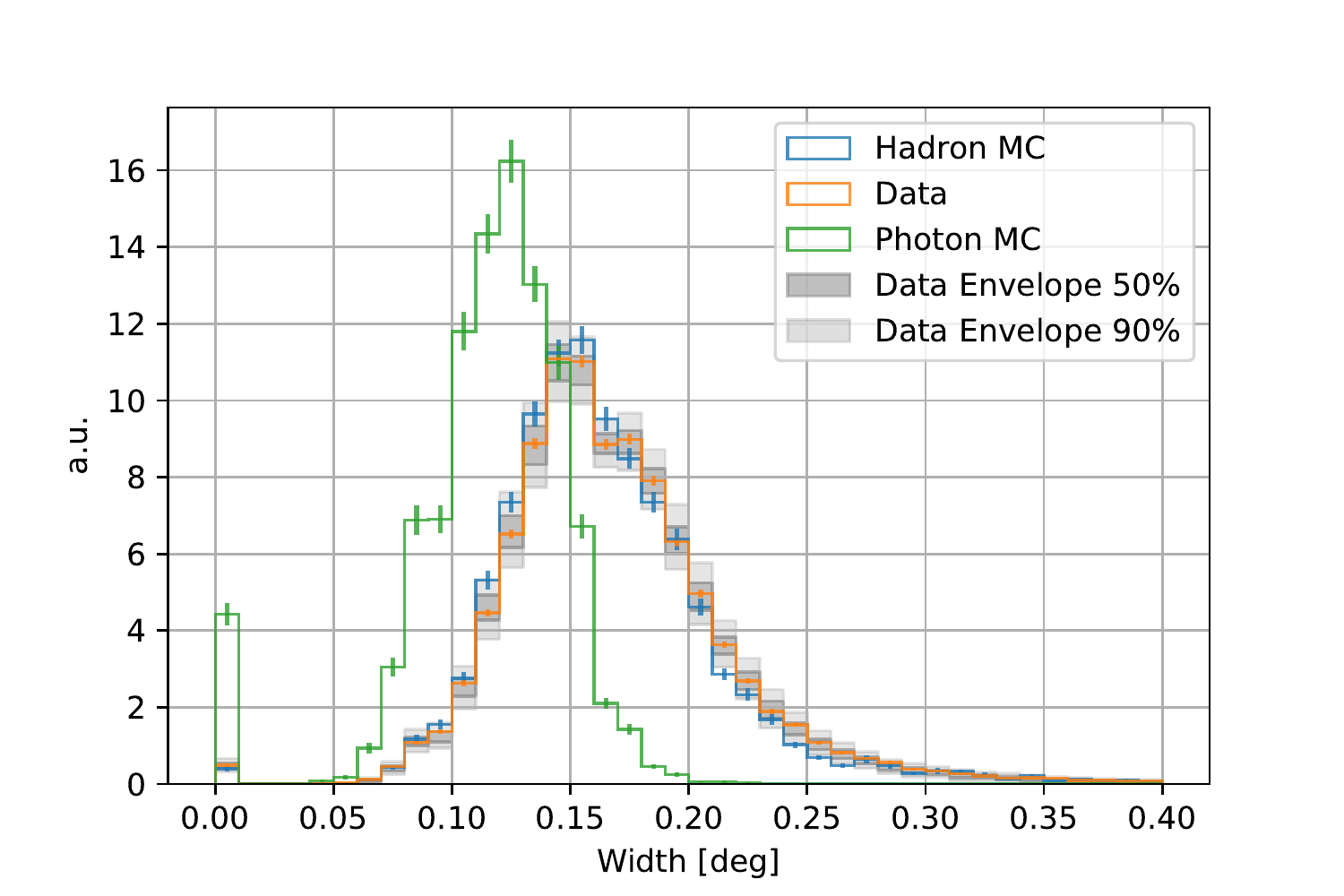}
    \caption{The image {\it width} ({\it second moment} of the charge distribution of the image pixels) as measured using cosmic ray data (orange), and MC simulations of cosmic rays (blue), and gamma-rays (green). The fluctuations in data are indicated as grey shaded areas for the 50\% and 90\% data envelopes, {defined as 50\% and 90\% containment range of the event-to-event fluctuations}.}
    \label{fig:width}
\end{figure}

\begin{figure}
        \includegraphics[width=\columnwidth]{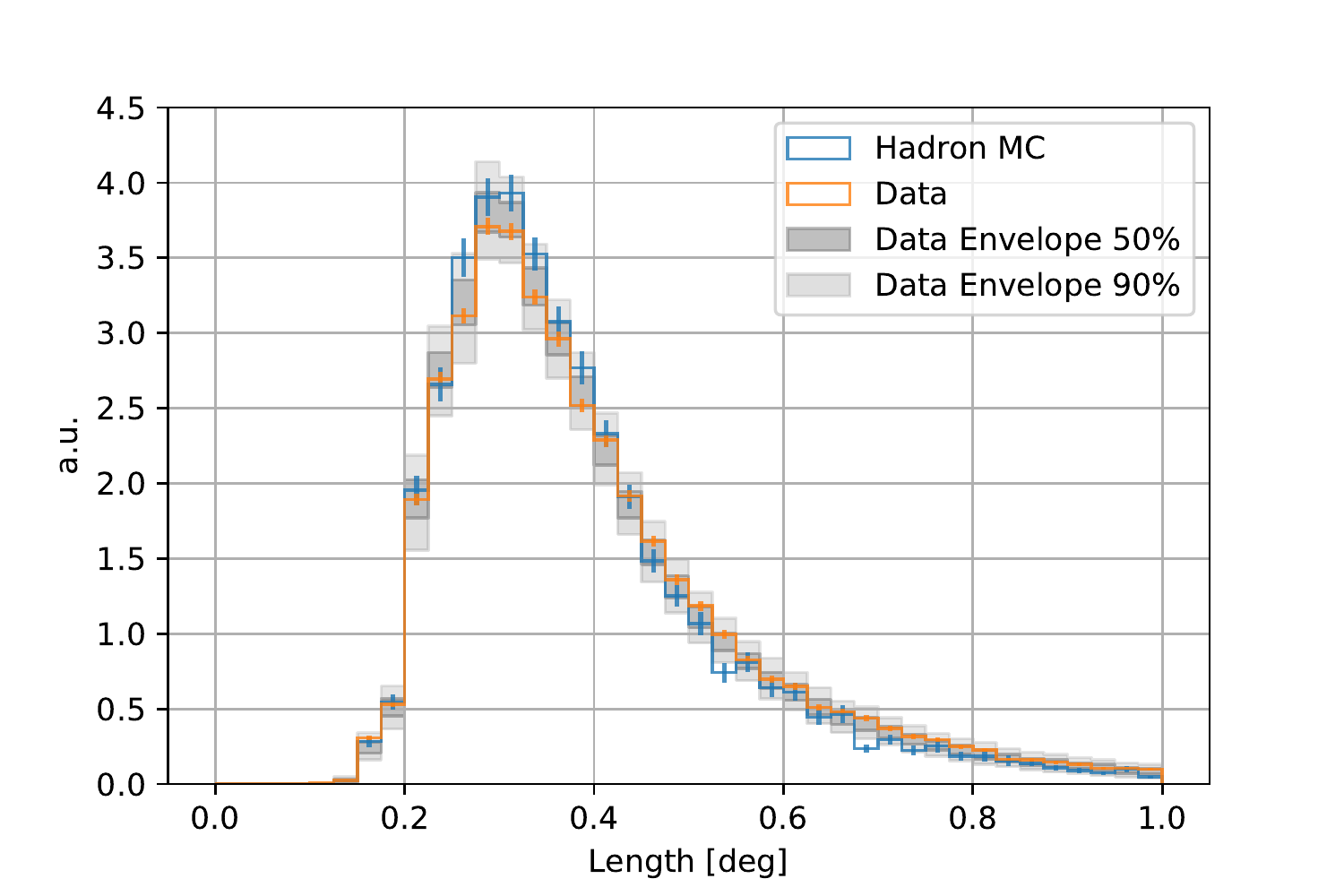}
    \caption{The image parameter {\it length} as measured using cosmic ray data (orange), and from MC simulations of cosmic rays (blue). Fluctuations from real data are shown in grey (50\% and 90\% envelopes, see caption of Fig~\ref{fig:width}).}
    \label{fig:length}
\end{figure}

\begin{figure}
        \includegraphics[width=\columnwidth]{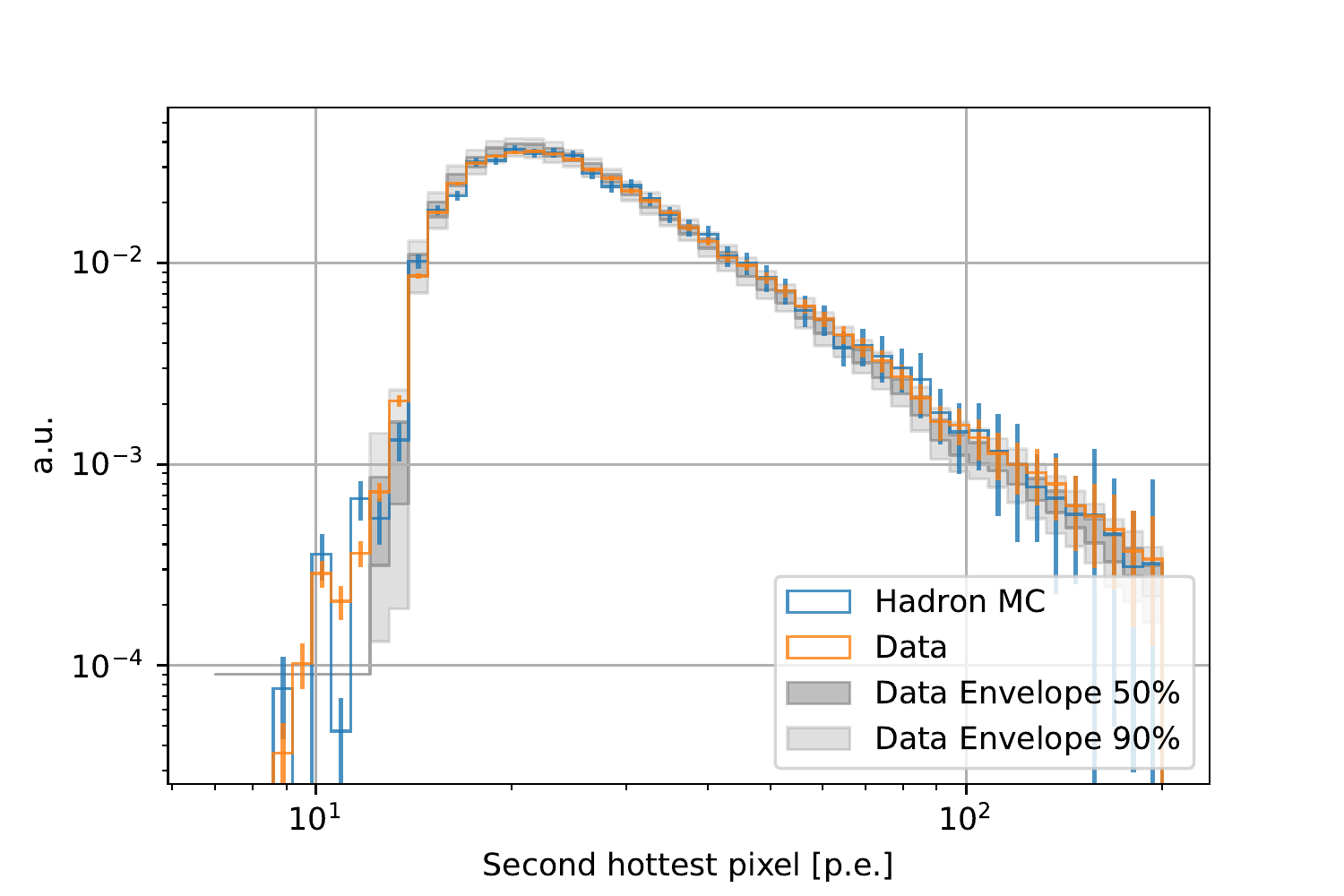}
    \caption{Comparison of the amplitude of the second-hottest image pixel. Cosmic ray data are shown in orange with fluctuations indicated by the grey shaded areas. MC simulations (blue) are weighted according to parametrisations of cosmic ray abundance are shown in blue.
    The grey shaded areas indicate data fluctuations as described in the caption of Fig~\ref{fig:width}.
   }
    \label{fig:secondhottestpixel}
\end{figure}

The comparisons show good agreement between real data and MC. Figure~\ref{fig:width}
shows the image parameter {\it width}, obtained from the {\it second moment} of the charge distribution in the image pixels, from background data of the first TAIGA-IACT, compared to the simulated {\it width} for hadrons and gamma rays.
The distributions for the parameter {\it length} for MC and cosmic ray data are shown in Figure~\ref{fig:length}.
The second brightest pixel in an air shower image is a robust low-level comparison parameter.
Figure~\ref{fig:secondhottestpixel}, shows that the simulations achieve a good reproduction of this distribution, except at the highest amplitude values, where the limited simulated energy range leads to a depletion in event statistics.

This fact tells us that the amplitude scale is well understood for the first TAIGA IACT. With the two pixel trigger, and a lower cleaning level well below the threshold, the peak {\bf${A_p}$} of the second-hottest pixel distribution without cuts (not shown here) directly describes the single pixel trigger thresholds. Gaussian fits to these peaks for different observation days throughout the season result in mean values of $9.7\,\mathrm{p.e.}$ to $11.3\,\mathrm{p.e.}$ with standard deviations of about $30\%$ and show a slight systematic variation. This indicates a stable operation during the season.
{The energy resolution of TAIGA was found to be 25\% in Monte Carlo simulations.
From uncertainties on the quantum and photo electron collection efficiencies of PMTs, the mirror reflectivity, the atmosphere, the light guide and plexiglas transmissions, we estimate an uncertainty on the energy scale of 20\%. 
Interestingly, \citep{2022arXiv220311502D} have observed that the systematic scaling factors required to accommodate a comprehensive collection of VHE results from different experiments was ranging from 0.89 to 1.14.
An uncertainty on the energy scale of 20\% translates into a systematic uncertainty on the flux of about 35\%.}

Further comparisons using low-level (pixel multiplicities, size distribution) or high-level parameters (higher image moments) also resulted in good agreement between the MC simulation and real data.

Some issues of the camera design described in Section~\ref{sec:iact1} will be addressed in future TAIGA IACT cameras. The above-mentioned thick "walls" of the Winston cones together with the thick Plexiglas entrance window reduce the amount of light collected from an air shower and increases the energy threshold of IACT1.
The large pixel size and the comparatively long charge integration time of the MAROC slow shaper result in a higher level of LoNS integration. Finally, further contributions to an increased energy threshold of the instrument for the Crab Nebula observations are the low altitude of the TAIGA observation site at $\thicksim$675\,m above sea level, and the latitude of the TAIGA experiment, resulting in a culmination of the Crab Nebula at a zenith angle of about 29$^\circ$. 
%

\begin{figure}
        \includegraphics[width=\columnwidth]{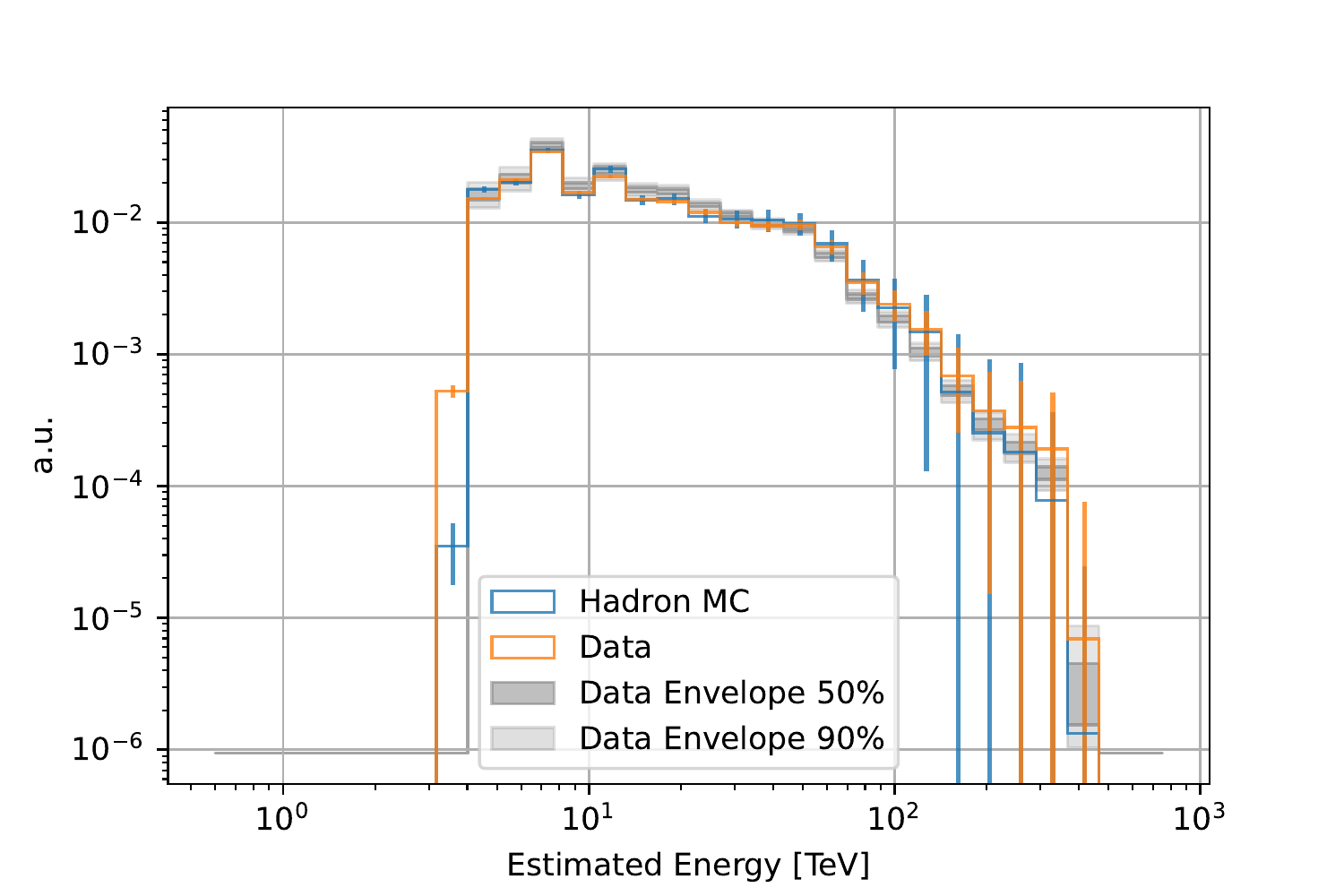}
 \caption{Expected differential cosmic ray event rate in arbitrary units from MC (blue) and real data (orange) as a function of reconstructed energy. Both distributions are normalized and are in good agreement. Fluctuations in real data are indicated by grey shaded areas. The expected rate distribution was obtained from weighting effective areas for proton, He, N, and Fe with a parametrisation of the cosmic ray spectra for all particles. Using gamma-ray simulations (not shown) yields an energy threshold of \ethreshold.
 Please note: this figure shows the energy distribution of hadrons as reconstructed using the random-forest energy estimator for gamma-rays. No effective areas for hadrons are taken into account. Therefore, this is not an energy spectrum of primary cosmic rays.
    The grey shaded areas indicate data fluctuations as described in the caption of Fig~\ref{fig:width}.
 }
    \label{fig:rate}
\end{figure}

The expected differential rate of cosmic rays was calculated from effective areas for protons, helium, nitrogen, and iron, weighed according to the parameterisation of cosmic ray spectra mentioned above.
The resulting expected differential all-particle cosmic ray rate is shown in Figure~\ref{fig:rate}. The predicted hadron trigger rate matches the measured trigger rate well.

\begin{figure}
        \includegraphics[width=\columnwidth]{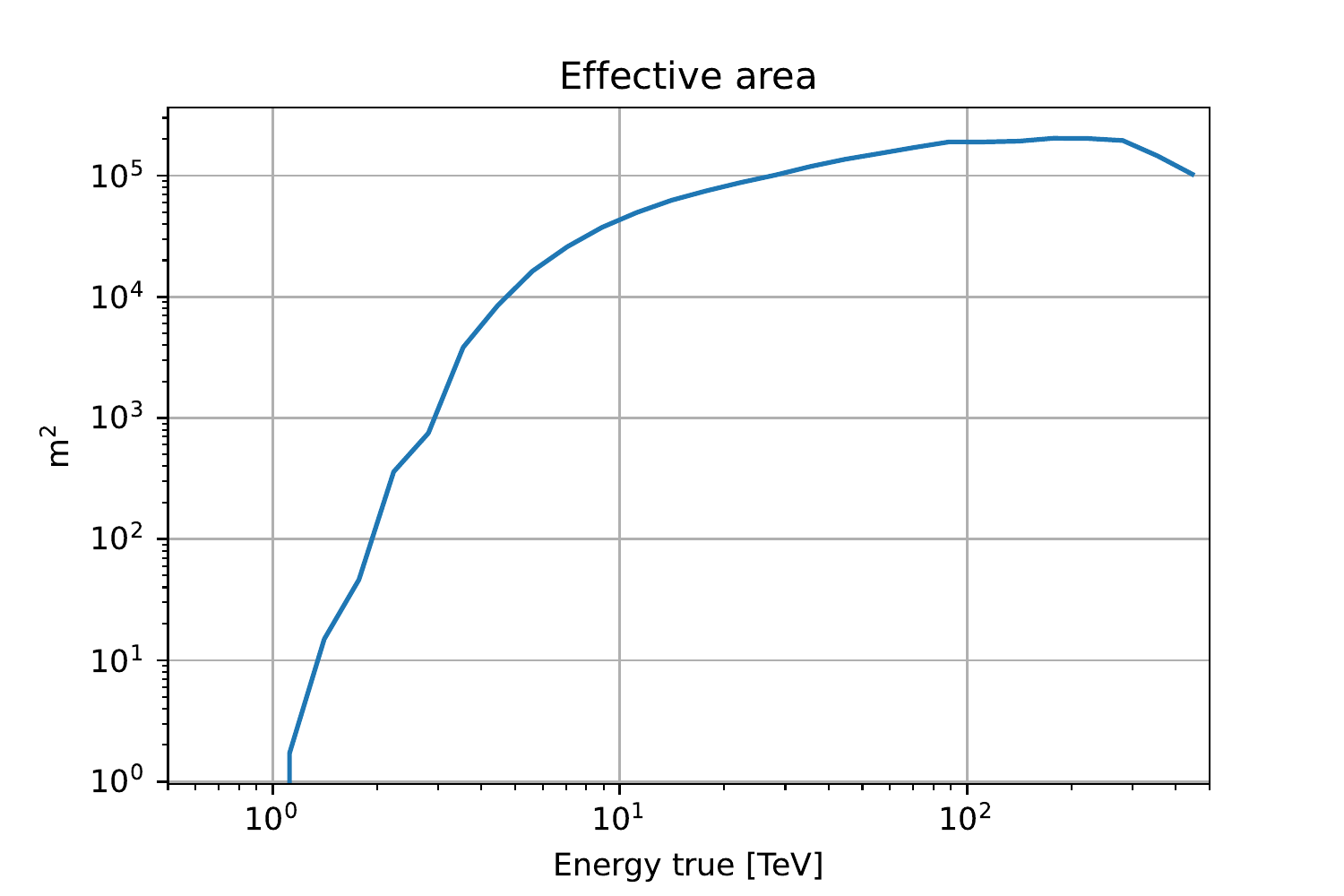}
    \caption{
    Effective area of the first TAIGA-IACT for the detection of gamma-rays using the random forest analysis described in the text. This effective area was calculated taking into account the observation conditions of the present Crab Nebula data set, and represents the average for all observations.}
    \label{fig:aeff}
\end{figure}

The effective area for gamma rays from MC simulations is shown in Figure~\ref{fig:aeff}. It was obtained using the random forest analysis described above. Effective areas A$_\mathrm{eff}\mathrm{(E,z)}$, depending on the energy E, and the zenith angle z, were calculated for each individual observation and averaged over time. The effective area shown is the average for all observations. 
As can be seen in this figure, the effective area in the energy range from 50\,TeV to 300\,TeV corresponds to an effective radius of the order of 220-250\,m around the IACT.

%
Given that the simulations describe the detector well, the gamma-ray effective area obtained from simulations can be used to calculate the energy threshold of the first IACT for gamma-rays.
The effective area for the current instrument design after reconstruction cuts 
is weighted
with a Crab Nebula-like spectrum. The peak of the differential energy distribution obtained is defined as the energy threshold and was found to be \ethreshold~in MC-energy.



\section{Crab Nebula Observations}
\subsection{Real data selection}
\label{sec:realdata}
In total, 204\,h of observations of the Crab Nebula pulsar were taken during the 2019/2020 season. After dead-time correction, 180\,h remain\footnote{The dead-time is estimated from the minimum time difference of consecutive events.}. Observations were carried out in wobble mode, using alternating offsets of the pointing positions of 1.2$^\circ$ to the position of the 
Crab Nebula pulsar. With a field of view of 9.6$^\circ$, a larger wobble radius would be possible if the trigger sensitive area were not limited to 3.7$^\circ$ (see Section~\ref{sec:iact1}).
Data are structured in portions of 2 minutes. Between portions, the anode currents of each PMT are measured.
The wobble direction is alternated between longer data segments, defined as observations (adopting the terminology of gammapy).
The selection criterion for good quality data is a cut on the trigger rate of a portion, rejecting portions with rates {lower or} greater than an expected zenith-angle corrected rate by more than two standard deviations. (7.35\,Hz to 11.96\,Hz)
These rates are calculated after applying a cut in the image size of {\it size}$>$80\,p.e..
Furthermore,
only portions pointing at altitudes above 56$^\circ$ are kept.
%
Applying these data quality cuts, in total \livetime of live-time corrected observations on the Crab Nebula remain. {During cold winter nights the observation conditions are good. However, because of the comparatively low observation height, adverse weather conditions are frequent during the rest of the year in the Tunka valley. Losses of up to 50\% are typical for an observation season. Further losses are due to current monitoring measurements and dead time (20\%).}


\subsection{Data calibration}
\label{sec:cal}
The pedestal value of each pixel is measured via the mean value of a Gaussian fit around the peak of the pixel amplitude distribution for each set of 100k events.  The pedestal value obtained is subtracted from the pixel amplitude. Image pixels with too wide pedestal distributions (e.g. due to bright stars such as Zeta Tauri) are rejected from further image analysis.
The raw pixel amplitudes measured in digitization units of Volts,
are converted to photo electrons using the F-factor method
\citep{1997ICRC....7..265M}
which
describes the multiplication process of electrons by the PMT dynode system.
For the purpose of calibration, the PMTs are illuminated by a mounted on the reflector dish LED system.
Additional standard calibration steps are
the pedestal subtraction and the mapping of low- to high-gain channels to a common photoelectron scale.
The calibrated images are compared to the simulated images in Section~\ref{sec:sim}, which demonstrates good agreement between real data and Monte Carlo for
image width (Figure~\ref{fig:width}),
image length (Figure~\ref{fig:length}),
amplitude of the second hottest pixel (Figure~\ref{fig:secondhottestpixel}),
and differential cosmic ray event rate (Figure~\ref{fig:rate}).

%
\subsection{Results: signal and energy spectrum of the Crab Nebula}
\label{sec:results}
%
%
%
The random forest-based reconstruction in combination with a high-level gammapy analysis developed and tested with simulations (Section~\ref{sec:sim}) is applied to calibrated (Section~\ref{sec:cal}) and quality selected (Section~\ref{sec:realdata}) data.
%
%

For high-level analysis, i.e. for the calculation of the significance of the excess and for the reconstruction of the spectrum, the gamma-py package is used \citep{gammapy:2017,axel_donath_2021_5721467}.

The excess from the direction of the Crab Nebula can be represented using an event-by-event accumulation of a histogram of
the squared angular distance $\theta^2$, defined as the angle between the vector pointing to a reference position (e.g. source position) and the vector pointing in the reconstructed event direction. A $\theta^2$ histogram with equidistant bin division covers an almost equal solid angle per bin within the range $\theta^2$ considered here.

Such a histogram is made for the on-source region, taking the position of the Crab Nebula pulsar as the reference position, and for different off-source pointings, used for the estimation of the background. The off-source positions are located along an arc at the same distance from the camera centre as the on-source position, therewith assuring the same camera acceptance as the on-source region (mirrored background model).

\begin{figure}
\includegraphics[width=\columnwidth]{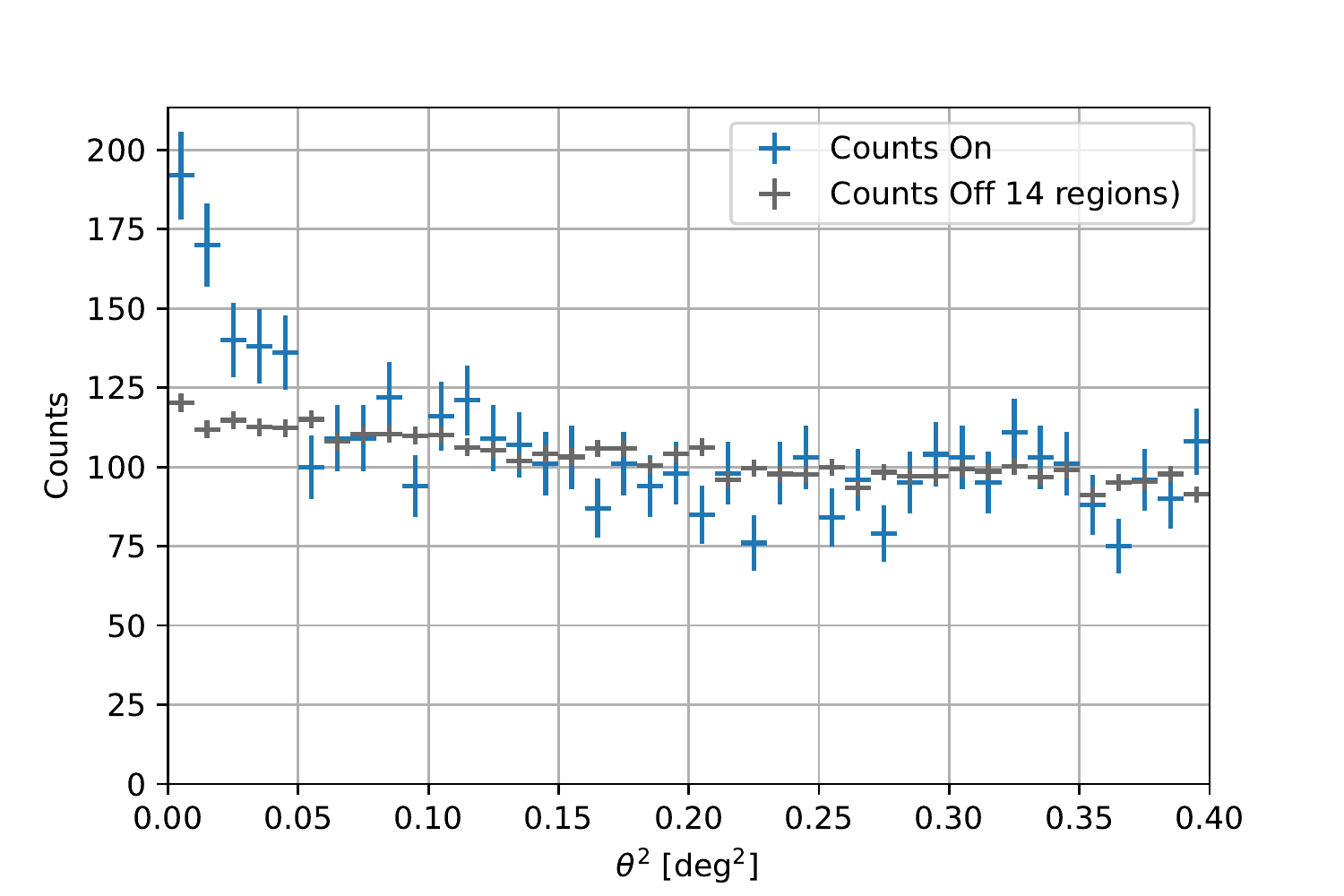}
\caption{
\label{fig:thetasq-crab}
Distribution of the square of the angular distance $\theta^2$, between the reconstructed arrival direction of an event and the direction of the Crab Nebula. Distributions are shown for the on-source region (blue), and an average distribution of \noffregions~off-source regions (dark grey). The background estimate from different off-regions is uncorrelated within the signal region. With increasing values of $\theta^2$ the off regions start overlapping. Therefore, the error bars are underestimated at large angular distances but not inside the signal region.
}
\end{figure}

The resulting histograms are shown in Figure~\ref{fig:thetasq-crab}. The blue crosses represent the on-source count statistics with statistical error bars. The grey crosses show the corresponding background distribution from off-regions.
For a reduction of the statistical uncertainty on the estimation of the number of background events, in total, \noffregions~off-source regions are used. Given that the on- and off-source regions are located at the same distance from the camera centre (wobble radius), the acceptance ratio $\alpha$ between the on-source and off-source estimates is $\alpha = 1/\noffregions = 0.0714$.

The on- and off-source distributions of $\theta^2$ are consistent for angular distances $\theta^2>0.1^\circ$. Below that value, the on-source distribution shows an excess towards the direction of the Crab Nebula (i.e. near $\theta^2=0$), while the off-source histogram essentially remains flat.

Events are selected, using a cut on $\theta^2<0.0324$deg$^2$, resulting in event counts for on-source (N$_\mathrm{on}$) and off-source (N$_\mathrm{off}$) events respectively. The cut value used corresponds to the 68\% containment radius of simulated events in the energy range from 20\,TeV to 100\,TeV.
{At the energy threshold \ethreshold, the 68\% containment radius is at 0.3\,deg, which improves at higher energies.}

Gammapy uses a Poissonian likelihood ratio approach to define a test statistics difference $\Delta TS$. As in \citep{lima}, the significance $S$ in terms of standard deviations $\sigma$ is then defined as $S=\sqrt{\Delta TS}$.
Application of the $\theta^2$ cut and constraining the reconstructed events to those estimated energy bins, where the effective area in true energy is larger than 5\% of the maximum effective area (gamma-py SafeMaskMaker) yields a total of N$_\mathrm{on}$\,=\,\non, N$_\mathrm{off}$\,=\,\noff, resulting in \excess~excess events, and a significance of \signi\,$\sigma$.

With an established signal from the direction of the Crab Nebula, 
the energy spectrum can be reconstructed in a next step. The random forest regressor introduced in Section~\ref{sec:reco} is used to reconstruct the energy for the on and off-source events surviving the $\theta^2$-cut.

\begin{figure}
\includegraphics[width=\columnwidth]{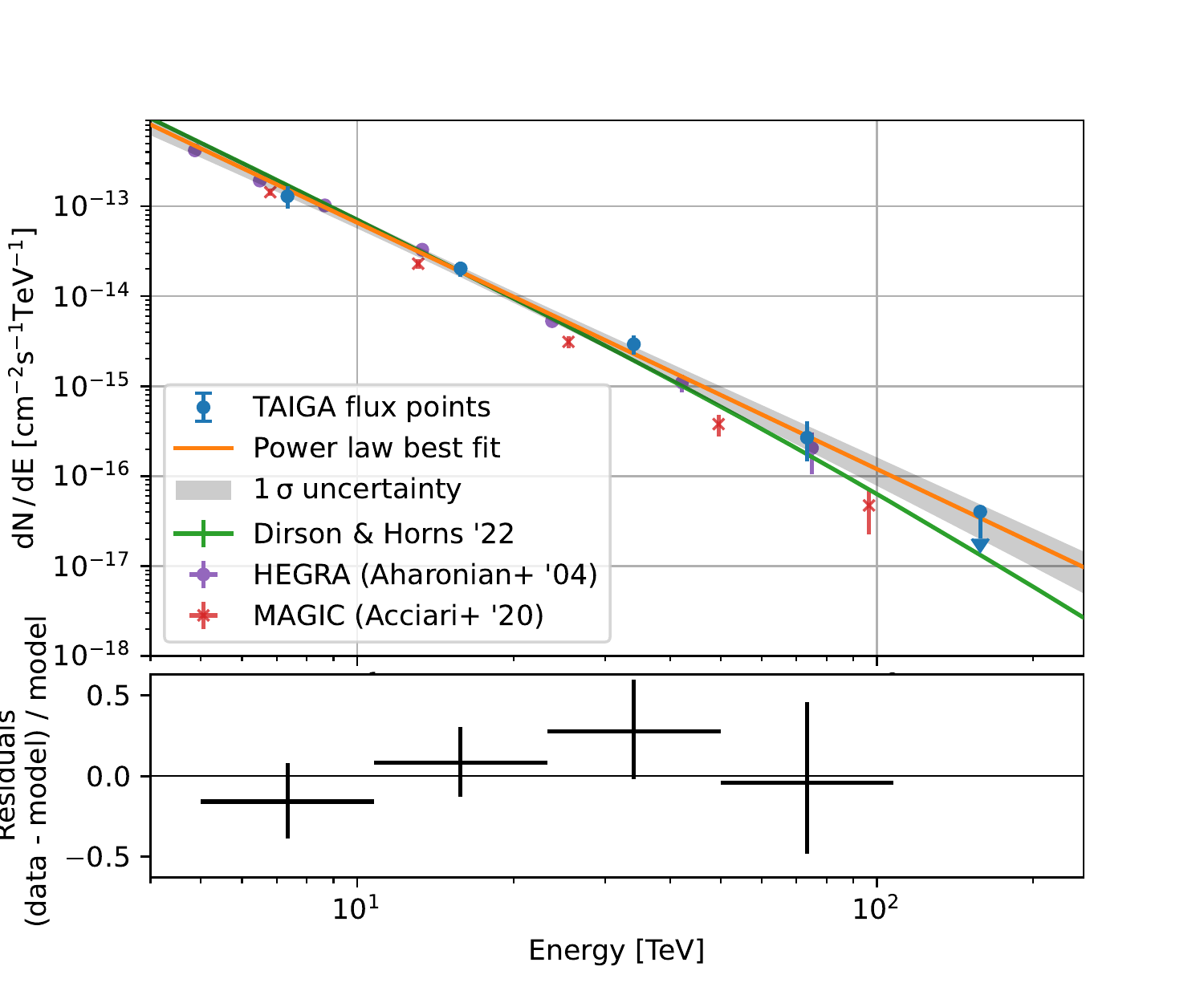}
\caption{
\label{fig:spectrum-crab}
The differential energy spectrum of the Crab Nebula from \livetime of TAIGA observations using one single wide field of view IACT. The applied forward-folding technique results in the grey shaded area, corresponding to the 68\,\% confidence interval of the best fit power law model (orange line). Flux points resulting from the spectral unfolding are shown in blue.
For comparison, a fit of  a model to a
compilation of several VHE data sets \citep{2022arXiv220311502D} is
overlaid to the data in green, and energy spectra from HEGRA and MAGIC are also shown.
The bottom part shows the residuals between TAIGA data and power law model.
}
\end{figure}

A forward-folding likelihood method provided by gamma-py is applied to reconstruct the energy spectrum and to calculate the flux in discrete energy bins.
The method requires the generation of instrument response functions (IRFs), which describe the point spread function (PSF), the effective area of the instrument, and the energy migration matrix.
The IRFs for TAIGA were specifically calculated for each observation. The conversion to a FITS format understood by gamma-py was done with pyirf \citep{maximilian_nothe_2022_5833284},
a python library developed for the generation of IRFs for CTA, but also applicable to other IACTs.
The pyirf package is used for normalising the PSF and for conversion to the FITS file format (GADF-FITS format).
One observation is defined as a real data interval with fixed celestial pointing of an average duration of 15 minutes. The IRFs were adapted to each observation using MC simulations matching the altitudes in real data ($\pm$2$^\circ$).
Assuming a power law spectral shape of the Crab Nebula of
\begin{equation}
    \frac{d\Phi}{dE} = \Phi_0 \left(\frac{E}{\eref}\right)^{-s}
    \label{eq:pl}
\end{equation}
the application of the forward-folding method results in a reconstructed energy spectrum shown in Figure~\ref{fig:spectrum-crab}.
The reference energy ($\eref$) was chosen close to the decorrelation energy (13.5\,TeV).

The resulting spectral shape is consistent with the power law assumption (Eq.~\ref{eq:pl}),
In the energy range from $\thicksim$5\,TeV to $\thicksim$100\,TeV. 
The best-fitting parameters were estimated as
$\Phi_0$\,=\,\phizero, and $s$\,=\,\spectralindex.
The statistical errors were obtained from the covariance matrix whose off-diagonal element is 0.07, indicating a weak correlation of the fit parameters.

The grey shaded area in Figure~\ref{fig:spectrum-crab} indicates statistical uncertainties only.
A log-parabola fit was also tested but not preferred
\footnote{The resulting test statistics difference of $\Delta TS=0.14$ between both models - assuming a $\chi^2$-distribution with one degree of freedom - corresponds to a p-value of 0.71.}. 
A summary of flux values per energy bin is given in Table~\ref{tab:especstats}.

The reconstructed spectrum is compatible to data from other experiments. For comparison, a fit of a model \citep{2022arXiv220311502D} to data from different experiments is shown in Figure~\ref{fig:spectrum-crab}. Furthermore, the energy spectra from MAGIC very large zenith angle observations \citep{2020A&A...635A.158M}, and from HEGRA \citep{2004ApJ...614..897A} are also shown.

\begin{table*}
    \centering
    \begin{tabular}{r|r|r|r|r|r|r|r|r}
    \hline
        E & E$_{1}$ & E$_{2}$       & Flux & Flux error positive & Flux error negative & $\sqrt{TS}$ & $n_{pred-excess}$ & counts  \\
        
        \multicolumn{3}{c}{TeV} & \multicolumn{3}{c}{$\cdot$10$^{-15}$TeV$^{-1}$cm$^{-2}$s$^{-1}$} & \multicolumn{3}{}{} \\ \hline
7.339 & 5.00   & 10.77 & 129.00 & 37.00 & 35.00 & 3.939 & 51.73   & 195 \\
15.810 & 10.77 & 23.21 & 20.30  & 4.10 & 3.90 & 5.843 & 64.00   & 155 \\
34.070 & 23.21 & 50.00  & 2.90   & 0.74 & 0.68 & 5.002 & 35.14   & 70 \\
73.390 & 50.00  & 107.70 & 0.27   & 0.14 & 0.12 & 2.368 & 10.64   & 26 \\
158.100 & 107.70 & 232.10 & $<$0.02  & - & -       & -0.296& -0.6357 & 4 \\

    \end{tabular}
    \caption{Fluxpoint values for the energy spectrum shown in Figure~\ref{fig:spectrum-crab}. The fluxes were determined via the Gammapy FluxPointEstimator using a forward folding fit under the assumption of a source with a power law spectrum with an index of 2.74. The obtained $\sqrt{TS}$ gives the significance of that bin. $n_{pred}$ is the number of expected events (excess + background) for that bin, $n_{pred-excess}$ is the number of expected excess events and counts is the number of events in the on region.}
    \label{tab:especstats}
\end{table*}

%
%
In an independent processing and analysis chain, a classical dynamical cut-based Hillas analysis was applied to a part of the data used here. It yielded a total of 162 excess events with a significance of 5.6\,$\sigma$ \citep{2021BRASP..85..398S}.
{The analysis is compatible with the results shown here. A thorough comparison 
will be published elsewhere \citep{kuzmichev2022}.}
%

\section{Conclusions and Outlook}

The first wide angle imaging air Cherenkov Telescope of the TAIGA experiment was operated in stand alone mode in the 2019/20 season.
Comparisons of MC simulations to real data show that the instrument
is well understood.
Even though using a comparatively large PMT pixel size of 0.36$^\circ$ (flat-to-flat), the TAIGA IACT, operated in monoscopic stand-alone mode, still provides a competitive angular resolution and gamma-hadron separation.
At the high energies which are at the focus of TAIGA, the increased LoNS caused by large pixels is less of an issue than at lower energies.

While different aspects of the first TAIGA IACT design can be improved
towards higher sensitivity and lower energy threshold, nevertheless 
using only one IACT with the current design provides a solid detection of the standard candle of gamma-ray astronomy under the extreme weather conditions in the Tunka valley in Siberia.
Observations of the Crab Nebula above \ethreshold~yield an excess of \signi\,$\sigma$ statistical significance {for the} \livetime of good quality observations. {The reconstructed energy spectrum} is compatible with a power law.
{The highest bin in the energy spectrum, extending from 50\,TeV to 108\,TeV shows a signal strength of 2.4\,$\sigma$.}

The broad-band measurement of the inverse Compton energy spectrum provides information on the magnetic field strength and its radial changes as has been pointed out recently \citep{2022arXiv220311502D}. Specifically, repeated measurements over several years will be essential to confirm the possibility that the magnetic field is dynamically evolving as suggested by the combination of measurements over the past two decades \citep{2022arXiv220311502D}. 
Statistics at the high energy end of the spectrum will grow with additional future data and additional telescopes being put into operation.

The IACT is planned to be operated in a hybrid array of 5 IACTs (3 built, 2 more planned) and 120 HiSCORE timing array stations.
Understanding the first IACT is an important step towards realizing the implementation of the new hybrid technique that will
combine the HiSCORE angle integrating air Cherenkov, timing detectors with imaging air Cherenkov telescopes.
Additionaly, scintillator detectors, both above and under ground, will allow the measurement of the muon component of air showers in future.
The expected sensitivity of the planned 5\,km$^2$ stage is better than 10$^{-13}$photons\,cm$^{-2}$s$^{-1}$ {\citep{budnev2020}}.

%

With these results, and previously published results on HiSCORE
(see, e.g.
\cite{Porelli2020TAIGA-HiSCORE:}) both air Cherenkov components of TAIGA, the TAIGA-HiSCORE, and the TAIGA-IACT detectors are operational, with a total instrumented area in excess of 1\,km$^2$. The main goal of the experiment is within reach: the hybrid reconstruction of gamma-ray sources in the energy range from TeV to several hundreds of TeV. 
%
The basic idea of the hybrid method is to use the strengths of the imaging and timing techniques. The HiSCORE timing array provides a good reconstruction of primary arrival direction, and EAS core position.
These parameters are combined with the imaging information of an IACT, resulting in a competitive
gamma-hadron separation using a single IACT combined with HiSCORE stations \citep{kunnas:phd}.
Using the hybrid method instead of the stereoscopic approach, will allow to fully exploit the large effective area of each single IACT (see Figure~\ref{fig:aeff}), with high reconstruction quality at core impact distances of 200-300\,m.
%
Consequentially, when using the hybrid method which is currently under development, we expect to be able to cover up to 1\,km$^2$ with 4 IACTs and $\thicksim$120 HiSCORE stations in the important energy range from several 10s to few 100s of TeV.


%
%

\section*{Acknowledgements}

We thank all the members of the TAIGA collaboration, for providing an excellent site infrastructure, years of experimental build-up, raw data calibration, and valuable discussions.
We acknowledge the support of the UNU “Astrophysical Complex of MSU-ISU” (agreement EB-075-15-2021-675).
We acknowledge Dieter Horns for his active support of the project at all stages of its development, from hardware contributions to the software analysis and the discussion of results and publications.
We thank Dirk Ryckbosch from the University of Gent for a valuable contribution of PMTs to the project.
We acknowledge Christian Spiering for organizing PMTs for the 3 imaging cameras of the TAIGA telescopes as well as his steady support of the project.
This analysis was supported by the Deutsche Forschungsgemeinschaft (DFG, TL 51/6-1)
and by the Helmholtz Association (HRJRG-303) and by European Union’s Horizon 2020 programme (No.653477). M.T, A.K.A. and M.B. acknowledge support by the Deutsche Forschungsgemeinschaft (DFG, German Research Foundation)
under Germany’s Excellence Strategy — EXC 2121 “Quantum Universe” — 390833306.
\section*{Data Availability}

Raw data and calibration files were generated within the TAIGA experiment, and are not available. Reconstructed data were generated at University of Hamburg. Derived data, i.e.
the results of the spectral reconstruction are available in this paper, Table~\ref{tab:especstats}.



\bibliographystyle{mnras}
\bibliography{TAIGACrabNebula} 

\begin{thebibliography}{}
\makeatletter
\relax
\def\mn@urlcharsother{\let\do\@makeother \do\$\do\&\do\#\do\^\do\_\do\%\do\~}
\def\mn@doi{\begingroup\mn@urlcharsother \@ifnextchar [ {\mn@doi@}
  {\mn@doi@[]}}
\def\mn@doi@[#1]#2{\def\@tempa{#1}\ifx\@tempa\@empty \href
  {http://dx.doi.org/#2} {doi:#2}\else \href {http://dx.doi.org/#2} {#1}\fi
  \endgroup}
\def\mn@eprint#1#2{\mn@eprint@#1:#2::\@nil}
\def\mn@eprint@arXiv#1{\href {http://arxiv.org/abs/#1} {{\tt arXiv:#1}}}
\def\mn@eprint@dblp#1{\href {http://dblp.uni-trier.de/rec/bibtex/#1.xml}
  {dblp:#1}}
\def\mn@eprint@#1:#2:#3:#4\@nil{\def\@tempa {#1}\def\@tempb {#2}\def\@tempc
  {#3}\ifx \@tempc \@empty \let \@tempc \@tempb \let \@tempb \@tempa \fi \ifx
  \@tempb \@empty \def\@tempb {arXiv}\fi \@ifundefined
  {mn@eprint@\@tempb}{\@tempb:\@tempc}{\expandafter \expandafter \csname
  mn@eprint@\@tempb\endcsname \expandafter{\@tempc}}}

\bibitem[\protect\citeauthoryear{Aartsen et~al.}{Aartsen
  et~al.}{2013}]{IceCube:2013cdw}
Aartsen M.~G.,  et~al., 2013, \mn@doi [Phys. Rev. Lett.]
  {10.1103/PhysRevLett.111.021103}, 111, 021103

\bibitem[\protect\citeauthoryear{{Acciari} et~al.,}{{Acciari}
  et~al.}{2020}]{2020A&A...635A.158M}
{Acciari} V.~A.,  et~al., 2020, \mn@doi [\aap] {10.1051/0004-6361/201936899},
  \href {https://ui.adsabs.harvard.edu/abs/2020A&A...635A.158M} {635, A158}

\bibitem[\protect\citeauthoryear{{Aharonian} et~al.,}{{Aharonian}
  et~al.}{2004}]{2004ApJ...614..897A}
{Aharonian} F.,  et~al., 2004, \mn@doi [\apj] {10.1086/423931}, \href
  {https://ui.adsabs.harvard.edu/abs/2004ApJ...614..897A} {614, 897}

\bibitem[\protect\citeauthoryear{{Albert}, {Aliu}, {Anderhub}, {Antoranz}
  et~al.}{{Albert} et~al.}{2008}]{2008NIMPA.588..424A}
{Albert} J.,  {Aliu} E.,  {Anderhub} H.,  {Antoranz} P.,   et~al., 2008,
  \mn@doi [Nuclear Instruments and Methods in Physics Research A]
  {10.1016/j.nima.2007.11.068}, \href
  {https://ui.adsabs.harvard.edu/abs/2008NIMPA.588..424A} {588, 424}

\bibitem[\protect\citeauthoryear{Alemanno et~al.,}{Alemanno
  et~al.}{2021}]{PhysRevLett.126.201102}
Alemanno F.,  et~al., 2021, \mn@doi [Phys. Rev. Lett.]
  {10.1103/PhysRevLett.126.201102}, 126, 201102

\bibitem[\protect\citeauthoryear{{An} et~al.,}{{An}
  et~al.}{2019}]{2019SciA....5.3793A}
{An} Q.,  et~al., 2019, \mn@doi [Science Advances] {10.1126/sciadv.aax3793},
  \href {https://ui.adsabs.harvard.edu/abs/2019SciA....5.3793A} {5, eaax3793}

\bibitem[\protect\citeauthoryear{Astapov et~al.,}{Astapov
  et~al.}{2018}]{Astapov2018}
Astapov I.,  et~al., 2018, \mn@doi [Nuclear Instruments and Methods in Physics
  Research Section A: Accelerators, Spectrometers, Detectors and Associated
  Equipment] {10.1016/j.nima.2018.10.081}, 936

\bibitem[\protect\citeauthoryear{{Berezhnev} et~al.,}{{Berezhnev}
  et~al.}{2012}]{2012NIMPA.692...98B}
{Berezhnev} S.~F.,  et~al., 2012, NIMA, 692, 98

\bibitem[\protect\citeauthoryear{{Bernl{\"o}hr}}{{Bernl{\"o}hr}}{2008}]{2008APh....30..149B}
{Bernl{\"o}hr} K.,  2008, Astroparticle Physics, 30, 149

\bibitem[\protect\citeauthoryear{Blank, Tluczykont, Kuotb~Awad  et~al.}{Blank
  et~al.}{2021}]{Blank2021}
Blank M.,  Tluczykont M.,  Kuotb~Awad A.,   et~al., 2021, in 37th International
  Cosmic Ray Conference (ICRC2021).

\bibitem[\protect\citeauthoryear{Blin, Barrillon, Puzo, de~la Taille,
  Seguin-Moreau  et~al.}{Blin et~al.}{2008}]{blin:2008}
Blin S.,  Barrillon P.,  Puzo P.,  de~la Taille C.,  Seguin-Moreau N.,
  et~al., 2008, MAROC: Multi Anode Readout Chip., \url
  {http://hal.in2p3.fr/in2p3-00308906}

\bibitem[\protect\citeauthoryear{Borodin, Grebenyuk, Grinyuk, Pan, Sagan,
  Tkachev  \& Wischnevsky}{Borodin et~al.}{2020}]{borodin:2020}
Borodin A.,  Grebenyuk V.,  Grinyuk A.,  Pan A.,  Sagan Y.,  Tkachev L.,
  Wischnevsky R.,  2020, \mn@doi [Physics of Atomic Nuclei]
  {10.1134/S1063778820020076}, 83, 268

\bibitem[\protect\citeauthoryear{Breiman}{Breiman}{2001}]{Breiman:2001hzm}
Breiman L.,  2001, \mn@doi [Machine Learning] {10.1023/A:1010933404324}, 45, 5

\bibitem[\protect\citeauthoryear{{Budnev} et~al.,}{{Budnev}
  et~al.}{2020}]{budnev2020}
{Budnev} N.,  et~al., 2020, Phys. Atom. Nuclei, 83, 905–915

\bibitem[\protect\citeauthoryear{Cao et~al.,}{Cao
  et~al.}{2021}]{cao:lhaaso:2021}
Cao Z.,  et~al., 2021, \mn@doi [Nature] {10.1038/s41586-021-03498-z}, 594

\bibitem[\protect\citeauthoryear{{Cirelli}, {Panci}, {Petraki}, {Sala}  \&
  {Taoso}}{{Cirelli} et~al.}{2017}]{2017JCAP...05..036C}
{Cirelli} M.,  {Panci} P.,  {Petraki} K.,  {Sala} F.,   {Taoso} M.,  2017,
  \mn@doi [\jcap] {10.1088/1475-7516/2017/05/036}, \href
  {https://ui.adsabs.harvard.edu/abs/2017JCAP...05..036C} {2017, 036}

\bibitem[\protect\citeauthoryear{{Deil} et~al.,}{{Deil}
  et~al.}{2017}]{gammapy:2017}
{Deil} C.,  et~al., 2017, in 35th International Cosmic Ray Conference
  (ICRC2017). p.~766 (\mn@eprint {arXiv} {1709.01751})

\bibitem[\protect\citeauthoryear{{Dirson} \& {Horns}}{{Dirson} \&
  {Horns}}{2022}]{2022arXiv220311502D}
{Dirson} L.,  {Horns} D.,  2022, arXiv e-prints, \href
  {https://ui.adsabs.harvard.edu/abs/2022arXiv220311502D} {p. arXiv:2203.11502}

\bibitem[\protect\citeauthoryear{Donath et~al.,}{Donath
  et~al.}{2021}]{axel_donath_2021_5721467}
Donath A.,  et~al., 2021, gammapy/gammapy: v.0.19,
  \mn@doi{10.5281/zenodo.5721467}, \url
  {https://doi.org/10.5281/zenodo.5721467}

\bibitem[\protect\citeauthoryear{{Gabici} \& {Aharonian}}{{Gabici} \&
  {Aharonian}}{2007}]{2007ApJ...665L.131G}
{Gabici} S.,  {Aharonian} F.~A.,  2007, ApJL, 665, L131

\bibitem[\protect\citeauthoryear{{Heck}, {Knapp}, {Capdevielle}, {Schatz}  \&
  {Thouw}}{{Heck} et~al.}{1998}]{1998cmcc.book.....H}
{Heck} D.,  {Knapp} J.,  {Capdevielle} J.~N.,  {Schatz} G.,   {Thouw} T.,
  1998, {CORSIKA: a Monte Carlo code to simulate extensive air showers.}

\bibitem[\protect\citeauthoryear{Hengstebeck}{Hengstebeck}{2007}]{Hengstebeck2007Measurement}
Hengstebeck T.,  2007, PhD thesis, Humboldt-Universität zu Berlin,
  Mathematisch-Naturwissenschaftliche Fakultät I,
  \mn@doi{http://dx.doi.org/10.18452/15628}

\bibitem[\protect\citeauthoryear{{Hillas}}{{Hillas}}{1985}]{1985ICRC....3..445H}
{Hillas} A.~M.,  1985, ICRC, 3, 445

\bibitem[\protect\citeauthoryear{{H{\"o}randel}}{{H{\"o}randel}}{2003}]{2003APh....19..193H}
{H{\"o}randel} J.~R.,  2003, \mn@doi [Astroparticle Physics]
  {10.1016/S0927-6505(02)00198-6}, \href
  {https://ui.adsabs.harvard.edu/abs/2003APh....19..193H} {19, 193}

\bibitem[\protect\citeauthoryear{{Horns} \& {Jacholkowska}}{{Horns} \&
  {Jacholkowska}}{2016}]{2016CRPhy..17..632H}
{Horns} D.,  {Jacholkowska} A.,  2016, \mn@doi [Comptes Rendus Physique]
  {10.1016/j.crhy.2016.04.006}, \href
  {https://ui.adsabs.harvard.edu/abs/2016CRPhy..17..632H} {17, 632}

\bibitem[\protect\citeauthoryear{Kunnas}{Kunnas}{2017}]{kunnas:phd}
Kunnas M.,  2017, Studies of the performance of an IACT system for the TAIGA
  array

\bibitem[\protect\citeauthoryear{Kuzmichev et~al.}{Kuzmichev
  et~al.}{2021}]{kuzmichev2021}
Kuzmichev L.,  et~al., 2021, PoS(2021)ICRC

\bibitem[\protect\citeauthoryear{Kuzmichev, Budnev, Sveshnikova
  et~al.}{Kuzmichev et~al.}{2022}]{kuzmichev2022}
Kuzmichev L.,  Budnev N.,  Sveshnikova L.,   et~al., 2022

\bibitem[\protect\citeauthoryear{Li \& Ma}{Li \& Ma}{1983}]{lima}
Li T.,  Ma Y.,  1983, ApJ, 272, 317

\bibitem[\protect\citeauthoryear{{Lubsandorzhiev}}{{Lubsandorzhiev}}{2019}]{Lubsandorzhiev2019}
{Lubsandorzhiev} N.,  2019, in 36th International Cosmic Ray Conference
  (ICRC2019). p.~730

\bibitem[\protect\citeauthoryear{{Mirzoyan} \& {Lorenz}}{{Mirzoyan} \&
  {Lorenz}}{1997}]{1997ICRC....7..265M}
{Mirzoyan} R.,  {Lorenz} E.,  1997, in International Cosmic Ray Conference.
  p.~265

\bibitem[\protect\citeauthoryear{Mirzoyan, Kankanian, Krennrich
  et~al.}{Mirzoyan et~al.}{1994}]{Mirzoyan1994}
Mirzoyan R.,  Kankanian R.,  Krennrich F.,   et~al., 1994, Nucl. Instr. Meth.
  Phys. Res. A, 351, 513

\bibitem[\protect\citeauthoryear{N{\"o}the et~al.,}{N{\"o}the
  et~al.}{2022}]{maximilian_nothe_2022_5833284}
N{\"o}the M.,  et~al., 2022, cta-observatory/pyirf: v0.6.0 –2022-01-10,
  \mn@doi{10.5281/zenodo.5833284}, \url
  {https://doi.org/10.5281/zenodo.5833284}

\bibitem[\protect\citeauthoryear{Pedregosa et~al.,}{Pedregosa
  et~al.}{2011}]{scikit-learn}
Pedregosa F.,  et~al., 2011, Journal of Machine Learning Research, 12, 2825

\bibitem[\protect\citeauthoryear{Porelli}{Porelli}{2020}]{Porelli2020TAIGA-HiSCORE:}
Porelli A.,  2020, PhD thesis, Humboldt-Universität zu Berlin,
  Mathematisch-Naturwissenschaftliche Fakultät,
  \mn@doi{http://dx.doi.org/10.18452/21610}

\bibitem[\protect\citeauthoryear{Postnikov, Astapov, Bezyazeekov
  et~al.}{Postnikov et~al.}{2019}]{Postnikov2019}
Postnikov E.,  Astapov I.,  Bezyazeekov P.,   et~al., 2019, Experiment. Bull.
  Russ. Acad. Sci. Phys., 83, 955–958

\bibitem[\protect\citeauthoryear{{Schweizer}}{{Schweizer}}{2002}]{2002PhDT.......391S}
{Schweizer} T.,  2002, PhD thesis, Autonomous University of Barcelona, Spain

\bibitem[\protect\citeauthoryear{{Sveshnikova} et~al.,}{{Sveshnikova}
  et~al.}{2021}]{2021BRASP..85..398S}
{Sveshnikova} L.~G.,  et~al., 2021, \mn@doi [Bulletin of the Russian Academy of
  Sciences, Physics] {10.3103/S1062873821040365}, \href
  {https://ui.adsabs.harvard.edu/abs/2021BRASP..85..398S} {85, 398}

\bibitem[\protect\citeauthoryear{Yashin et~al.}{Yashin
  et~al.}{2015}]{Yashin2016}
Yashin I.,  et~al., 2015, in 34th International Cosmic Ray Conference
  (ICRC2015). p.~986

\bibitem[\protect\citeauthoryear{{Zhurov}}{{Zhurov}}{2019}]{2019ICRC...36..833Z}
{Zhurov} D.,  2019, in 36th International Cosmic Ray Conference (ICRC2019).
  p.~833

\bibitem[\protect\citeauthoryear{{Zhurov} et~al.}{{Zhurov}
  et~al.}{2021}]{Zhurov2021}
{Zhurov} D.,  et~al., 2021.

\makeatother
\end{thebibliography}






\bsp	
\label{lastpage}
\end{document}